\documentclass[a4paper,12pt]{article}

\usepackage{jheppub} 

\usepackage[T1]{fontenc} 
\usepackage{physics}
\usepackage{subcaption}

\usepackage{amsmath}
\usepackage{amsthm}
\usepackage{float}
\usepackage{braket}
\theoremstyle{definition}
\newtheorem{thm}{Theorem}[section]
\newtheorem*{thm*}{Theorem}
\newtheorem{defn}[thm]{Definition}
\newtheorem*{defn*}{Definition}
\newtheorem{lem}[thm]{Lemma}
\newtheorem*{lem*}{Lemma}

\newtheorem*{rem*}{Remark}

\newtheorem*{con*}{Conjecture}

\newtheorem*{cor*}{Corollary}
\newtheorem{prop}[thm]{Proposition}
\newtheorem*{prop*}{Proposition}

\newtheorem*{hypoth*}{Hypothesis}

\newtheorem*{claim*}{Claim}
\newtheorem*{prf}{Proof}

\newcommand{\fix}[1]{\textcolor{red}{[#1]}}

\title{Algebra of Hyperbolic Band Theory under Magnetic Field}

\author[a,b]{Kazuki Ikeda,}
\author[c]{Yoshiyuki Matsuki,}
\author[c]{Shoto Aoki}

\affiliation[a]{Co-design Center for Quantum Advantage $\&$ Center for Nuclear Theory, Department of Physics and Astronomy, Stony Brook University, Stony Brook, New York 11794-3800, USA}
\affiliation[b]{Department of Mathematics and Statistics
$\&$ Centre for Quantum Topology and Its Applications (quanTA), University of Saskatchewan, Saskatoon, Saskatchewan S7N 5E6, Canada}
\affiliation[c]{Department of Physics, Osaka University, Toyonaka, Osaka 5600043, Japan}

\emailAdd{kazuki7131@gmail.com}

\date{\empty}
\arxivnumber{\empty}

\abstract{We explore algebras associated with the hyperbolic band theory under a magnetic field for the first time. We define the magnetic Fuchsian group associated with a higher genus Riemann surface. By imposing the magnetic boundary conditions for the hyperbolic Bloch states, we construct the hyperbolic magnetic Bloch states and investigate their energy spectrum. We give a connection between such magnetic Bloch states and automorphic forms. Our theory is a general extension of the conventional algebra associated with the band theory defined on a Euclidean lattice/space into that of the band theory on a general hyperbolic lattice/Riemann surface.}

\begin{document} 
\maketitle 

\section{Introduction and Summary}
Up to now, band theory has made great strides in revealing the physical properties of solids in Euclidean space. On the other hand, constructing a theory of solid materials in curved space remains a remaining challenge. 2d closed surfaces with constant curvature can be classified as sphere, torus and Riemann surfaces with genus $g$ greater than 1~($g\ge2$). Of these, physics on the torus is the most well-studied in modern band theory, and a spherical surface is also preferred. On the other hand, condensed matter physics on Riemann surfaces with negative curvature, whose genus is greater than or equal to two has not received much attention. Here, negatively curved Riemann surfaces can be created by properly identifying the boundaries of the 2d hyperbolic surfaces. For example, a Riemann surface with $g=2$ is obtained by laminating the edges of the Poincar\'{e} disk represented by \{8,8\} tiles, which consists of 8 vertices each with 8 edges (Fig~\ref{fig:tiling}). Such structures are particularly well studied in algebraic geometry, and in relation to physics they play a very important role in string theory, where trajectories of the strings form Riemann surfaces.  More recently, hyperbolic extensions of the conventional energy band theories on the Euclidean space have attracted a general interest of authors, from various motivations~\cite{2020arXiv200805489M,2021arXiv210413314I,2021arXiv210501087B,2021arXiv210506490B,2022NatCo..13.2937Z,2022PhRvL.128p6402S,2022PhRvB.105x5301L,2022PhRvL.129x6402U,2023arXiv230111884I}. Those theories can be testable by a circuit quantum electrodynamics (cQED), which is a promising candidate of universal quantum computation~\cite{2019Natur.571...45K,2019PhRvR...1c3079J,2015JHEP...06..149P,2022NatCo..13.4373L}. Those hyperbolic extensions give us a new platform of material design, condensed matter, high energy physics, as well as mathematical physics. In particular, studying physical effects of the underlying space on electrons properties is crucial. For this purpose, introducing a magnetic field is the most common way for condensed matter physics. To this end, we aim at providing mathematical foundation of the hyperbolic band theory under a magnetic field. We formulate the tight-binding Hamiltonian on a Poincar'{e} tiling and address its algebra. Physical analysis and interpretations of our work are given in~ \cite{2021arXiv210413314I}. 

Our contributions to the hyperbolic band theory can be summarized as follows. Let $B$ be a magnetic field which is related with the gauge field as $A=\frac{B}{y}dx$. Then we define the \textit{magnetic} $PSL_2(\mathbb{R})$ as follows:

\textbf{Definition~\ref{def:magneticSPL}} We define a Lie group $PSL_2(\mathbb{R})[B]$ which is generated by 
\begin{align}
\begin{aligned}
        S&=(1+x^2-y^2)\pdv{}{x}+2xy\pdv{}{y}+2iBy\\
        T&=\pdv{}{x}\\
        U&=2x\pdv{}{x}+2y\pdv{}{y}.
\end{aligned}
\end{align}
We call it \textit{magnetic} $PSL_2(\mathbb{R})[B]$.

The following theorem is fundamental for the study of the Bloch condition under a magnetic field.

\textbf{Theorem~\ref{thm:magneticPSL}} If $B=1/q$, $PSL_2(\mathbb{R})$ is a $q$-fold covering of $PSL_2(\mathbb{R})[B]$.

Now let us extend the notion of the Fuchsian group under a magnetic field. This is an extension of algebra acting on the magnetic Brillouin zone in 2d torus in the context of the conventional band theory.

\textbf{Definition~\ref{def:Fuchsian}}
We define {\it magnetic Fuchsian group} by
\begin{align}
\begin{aligned}
    \hat{\gamma}_j^B=&\exp(\mu \qty(\hat{U}_B
    \cos(\frac{j-1}{4}\pi ) -(2\hat{T}_B-\hat{S}_B) \sin(\frac{j-1}{4}\pi)) )\\
    =&\exp( -\frac{j-1}{8}\pi \hat{S}_B) \exp( \mu \hat{U}_B) \exp( \frac{j-1}{8}\pi \hat{S}_B).
\end{aligned}
\end{align}

The following theorem provides an argument for the definition of the magnetic Fuchsian group as a proper extension of the conventional Fuchsian group. on Poincar\'e tiling without magnetic field. Especially the result agrees with the Gauss-Bonnet theorem on a hyperbolic surface.

\textbf{Theorem~\ref{thm:Fuchsian}}
Let $\Gamma$ be a Fuchsian group of the $\{4g,4g\}$ tiling. The generators $\hat{\gamma}^B_j\ (j=1,\cdots ,2g)$ of magnetic Fuchsian group corresponding to $\Gamma$ satisfy
\begin{align}
    \hat{\gamma}_{2g}^B \cdots \hat{\gamma}_2^B (\hat{\gamma}_1^B)^{-1}(\hat{\gamma}_{2g}^B)^{-1}\cdots (\hat{\gamma}_2^B)^{-1} \hat{\gamma}_1^B=e^{i4(g-1)\pi B}.
\end{align}
Note that $\phi=4(g-1)\pi B$ is equal to the magnetic flux through the fundamental domain.

Our paper is organized as follows. In the next section, we give a short review of hyperbolic surface and Poincar\'e tiling. In Section 3, we present our main results. In Sections 4 and 5, we apply our results to the Hofstadter problem on a hyperbolic lattice.

\section{Magnetic Field on Hyperbolic Surface}
\subsection{General Setup}
Based on the Bloch band theory on a hyperbolic lattice \cite{2020arXiv200805489M}, we consider the quantum Hall effect on a hyperbolic lattice in the presence of a magnetic field perpendicular to the system. Let $\mathbb{H}$ be the upper half-plane equipped with its usual Poincar\'{e} metric $ds^2=\frac{dx^2+dy^2}{y^2}$. Let $A\in\Omega^1(\mathbb{H})$ be a one-form on $\mathbb{H}$. Using the Laplace-Beltrami operator, we can write the Hamiltonian with a magnetic field as 
\begin{align}
\begin{aligned}
\label{eq:Hamiltonian}
H&=\frac{1}{2m}\frac{1}{\sqrt{g}}(p_\mu-A_\mu)\sqrt{g}g^{\mu\nu}(p_{\nu}-A_\nu)\\
&=\frac{1}{2m}y^2((p_x-A_x)^2+(p_y-A_y)^2),
\end{aligned}
\end{align}
where $p_\mu=-i\partial_\mu$. Since we are interested in the quantum Hall effect, we take the constant magnetic field that can be written as $dA=B\omega$, where $\omega=y^{-2}dx\wedge dy$ is the area element and $B\in\mathbb{R}$ is some fixed value. We use $A=\frac{B}{y}dx$, then the Hamiltonian becomes 
\begin{align}
\begin{aligned}
    H&=\frac{y^2}{2m}\left(\left(p_x-\frac{B}{y}\right)^2+p_y^2\right)\\
    &\label{eq:V}=-\Delta+i\frac{B}{m}y\partial_x+\frac{B^2}{2m}. 
\end{aligned}
\end{align}

With the metric $\lambda^2(z)dzd\bar{z}~(z=x+iy)$, then the Laplace-Beltrami operator is defined by 
\begin{equation}
    \Delta=\frac{4}{\lambda^2}\frac{\partial}{\partial z}\frac{\partial}{\partial \bar{z}}=\frac{1}{\lambda^2}\left(\frac{\partial^2}{\partial x^2}+\frac{\partial^2}{\partial y^2}\right).
\end{equation}
With $ds^2=\frac{dx^2+dy^2}{y^2}$, the curvature is $K=-1$.

\subsection{General Prescriptions for States in Hyperbolic Surface under Magnetic Field}
\begin{figure}[H]
    \centering
    \includegraphics[width=8cm, bb=0 0 400 400]{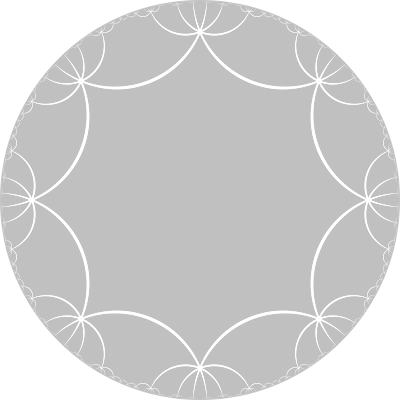}
    \caption{$\{8,8\}$-tiling of the Poincar'{e} disk.}
    \label{fig:tiling}
\end{figure}

In what follows, we show a general procedure to address electron states in a hyperbolic surface under a uniform magnetic field. Some concrete studies will be given in Sec. \ref{sec:MFG}, where detailed formulations of the magnetic Fuchsian group and Bloch states are given. We address electron states on a lattice, so-called a Poincar\'{e} tile~(Fig.\ref{fig:tiling}). 

\begin{enumerate}
    \item We propose the {\it magnetic Fuchsian group}, that commutes with the Hamiltonian~\eqref{eq:Hamiltonian} on the hyperbolic lattice. For this, we prepare the generators of  $SL_2(\mathbb{R})$ so that they commute with the Hamiltonian~\eqref{eq:Hamiltonian} and consider the magnetic translation.
    \item We construct the {\it magnetic hyperbolic Bloch state}, which is a Bloch-like state on a hyperbolic lattice. This can be done by imposing the magnetic Bloch condition. 
\end{enumerate}
Note that in a Euclidean lattice, the tight-binding Hamiltonian commutes with the translation operators, hence the Bloch states are their simultaneous eigenstates. In fact, the translation operators are commutative in the absence of a magnetic field. However, in a hyperbolic lattice, the Fuchsian group is non-commutative. Therefore the construction of the magnetic Bloch state is quite non-trivial. Nevertheless, we verify they exist and give their concrete form in Sec. \ref{sec:MFG}. 

\section{Magnetic Fuchsian Group}
\subsection{Review on the Fuchsian Group}
In this section, we give a brief review on some basic properties and mathematical background about a transformation group that acts on the hyperbolic plane. A discrete subgroup of the transformation group is called the Fuchsian group, and it determines the tiling of the hyperbolic plane.
\begin{defn}[Hyperbolic plane]
Let $\mathbb{H}$ be a Riemann manifold with a metric $ds^2=\frac{1}{y^2}(dx^2+dy^2)$ on upper half-plane $\Set{ z=x+iy \in \mathbb{C}| y>0 }$.
\end{defn}

Let $SL_2(\mathbb{R})$ be a special linear group defined by
\begin{align}
    SL_2(\mathbb{R})=\Set{ g= \left(\begin{array}{cc}
        a & b \\
        c & d
    \end{array} \right) |  \det g=1 }.
\end{align}
This group acts on the Hyperbolic plane $\mathbb{H}$ as follows:
\begin{align}
    g\cdot z=\frac{az+b}{cz+d} \ (g\in SL_2(\mathbb{R}),\ z\in \mathbb{H} ).
\end{align}
This action satisfies 
\begin{align}
\begin{aligned}
    g_1 \cdot ( g_2 \cdot z)=(g_1 g_2) \cdot z \\
    g\cdot z =(-g) \cdot z
\end{aligned}
\end{align}
for arbitrary $g_1,g_2,g \in SL_2(\mathbb{R})$ and $z \in \mathbb{H}$. Below, when there is no misunderstanding, "$\cdot$" indicating the action is omitted.

It is known that $sl_2 (\mathbb{R})$ is generated by the following three elements:
\begin{align}\label{eq:commutation relation of SL2}
    S=\mqty(0& 1 \\ -1 & 0),\ T=\mqty(0 & 1 \\ 0 & 0 ),\ U=\mqty(1 & 0 \\ 0 & -1 ),
\end{align}
where $S,T,$ and $U$ denote rotation, translation, and scaling, respectively. The commutation relations are given by
    \begin{align}
    [U,T]=2T,\ [U,S]=4T-2S,\ [S,T]=U.
\end{align}

The generator $S$ produces one-parameter family
\begin{align}
    e^{\theta S}=\mqty(\cos\theta & \sin \theta \\-\sin \theta &\cos \theta ) \ (\theta \in \mathbb{R})
\end{align}
in $SL_2(\mathbb{R})$. $e^{\theta S}$ acts on $z_0 \in \mathbb{H}$ as 
\begin{align}
    e^{\theta S} z_0=\frac{\cos \theta z_0 +\sin \theta}{ -\sin \theta z_0 +\cos\theta},
\end{align}
and we consider an orbit that is described by moving the parameter $\theta$.

\begin{prop}\label{prop:orbit of S}
For arbitrary $z_0 \in \mathbb{H}$, an orbit $\Set{e^{\theta S} z_0| \theta \in \mathbb{R} }$ is equal to
\begin{align}\label{eq:an orbit generated by S}
C=\Set{ z=x+iy \in \mathbb{H} | x^2+(y-a)^2=b^2},
\end{align}
where 
\begin{equation}
    a=\frac{x_0^2+y_0^2+1}{2y_0},\ b=\sqrt{a^2 -1}.
\end{equation}
Therefore the orbit generated by $S$ is a circle with center $z=ia$ and radius $b$. 
\end{prop}

\begin{prf}
Let $z(\theta)=e^{\theta S} z_0=x(\theta)+iy(\theta), z(0)=z_0$. To find a differential equation that $z(\theta)$ should satisfy, we differentiate $z(\theta)$ by $\theta$. Since $z(\theta)$ satisfies $z(\theta_1+\theta_2)=e^{\theta_1 S}z(\theta_2)$, it is sufficient to consider a derivative of $z(\theta)$ near $\theta=0$. From
\begin{align}
    \eval{\dv{}{\theta} z(\theta)}_{\theta=0}=1+x_0^2-y_0^2+i2x_0 y_0,
\end{align}
the equation to be satisfied by $x,y$ is
\begin{align}\label{eq:differential eq of z(theta)}
\begin{aligned}
    \dv{}{\theta}x&=1+x^2-y^2,\\
    \dv{}{\theta}y&=2x y.
\end{aligned}
\end{align}

For $x,y$ satisfying eq.~\eqref{eq:differential eq of z(theta)}, let $f(\theta)=x(\theta)^2+(y(\theta)-a)^2-b^2$ be a function of $\theta$, where $a,b$ are constants. Differentiating $f$ by $\theta$, we obtain
\begin{align}
    f^\prime(\theta)=2x(1+b^2-a^2+f).
\end{align}
Let $a^2-b^2=1$, $f \equiv0$ satisfies the above equation. Thus a closed loop defined by $x^2+(y-a)^2=b^2, a^2-b^2=1$ is the orbit of \eqref{eq:differential eq of z(theta)}. If the loop contains $z_0$, then
\begin{align}
    a=\frac{x_0^2+y_0^2+1}{2y_0},\ b=\sqrt{a^2 -1}.
\end{align}
\qed
\end{prf}

$T$ and $U$ generate 
\begin{align}
    e^{tT}=\mqty(1& t \\ 0 & 1 ),~e^{\mu U}=\mqty( e^{\mu} & 0 \\ 0 & e^{-\mu})
\end{align}
and they act on $z_0 \in \mathbb{H}$ as
\begin{align}
    e^{t T} z_0= z_0 +t,~e^{\mu U} z_0= e^{2\mu }z_0.
\end{align}

Any element $g \in SL_2(\mathbb{R})$ can be uniquely decomposed as
\begin{align}\label{eq:docomposition of SL2}
    g=e^{\theta S}e^{\mu U} e^{tT},
\end{align}
which is called the Iwasawa decomposition of the group.
Let $\Gamma$ be a discrete subgroup of $SL_2(\mathbb{R})$, so-called a Fuchsian group. Here discrete means that there is no element in $\Gamma$ near the identity element. For example, $SL_2(\mathbb{Z} )$ is a Fuchsian group, but $SL_2(\mathbb{Q})$ is not. 

The following Prop.~\ref{prop:def of fundamental domain} is well-known~\cite{katok1992fuchsian}.

\begin{prop}\label{prop:def of fundamental domain}
Let $\Gamma$ be a Fuchsian group of $SL_2(\mathbb{R})$. Then, there exists a subset $D$ in $\mathbb{H}$ that satisfy the following:
\begin{enumerate}
    \item $\mathbb{H}=\bigcup_{\gamma \in \Gamma} \gamma D$.
    \item If $\gamma \in \Gamma, \gamma \neq \pm 1$, then a measure of $D\cap \gamma D$ is zero.
    \item The edges of $D$ are geodesics. 
\end{enumerate}
Such a $D$ is called by a fundamental domain of $\Gamma$. 
\end{prop}

This proposition means that $\mathbb{H}$ can be divided by $D$. Here we can take $D$ to be a regular $p$-gon, and we can lay it out in $\mathbb{H}$ so that $q$ fundamental domains share one vertex. We call such a partition the $\{p,q\}$ tiling. 

Let 
\begin{align}
\begin{aligned}
    \gamma_j=&\mqty(\cos( \frac{j-1}{4g}\pi) & \sin( \frac{j-1}{4g}\pi) \\-\sin( \frac{j-1}{4g}\pi) & \cos( \frac{j-1}{4g} \pi) )
    \mqty (e^\mu & 0 \\ 0 & e^{-\mu}  )\mqty(\cos( \frac{j-1}{4g}\pi) & -\sin( \frac{j-1}{4g}\pi) \\\sin( \frac{j-1}{4g}\pi) & \cos( \frac{j-1}{4g} \pi) )\\
    =&\exp(\mu \qty(U \cos(\frac{j-1}{2g}\pi ) -(2T-S) \sin(\frac{j-1}{2g}\pi)) )
\end{aligned}
\end{align}
be generators of a Fuchsian group $\Gamma$, where $j=1,\cdots,2g$ and $\mu$ satisfies
\begin{equation}
    e^\mu=\frac{\cos \frac{\pi}{4g}}{\sin \frac{\pi}{4g}}+\sqrt{  \frac{ \cos^2 \frac{\pi}{4g}}{ \sin^2 \frac{\pi}{4g}} -1}.
\end{equation} 

This $\Gamma$ gives the $\{4g,4g\}$ tiling. The fundamental domain $D$ is a regular $4g$-gon, and the vertices $v_1,\cdots v_{4g}$ are located at
\begin{align}\label{eq:vertex of D}
\begin{aligned}
    v_{4g}&=e^\mu-\frac{\sin \frac{\pi}{4g} }{ \cos \frac{\pi}{4g}}+i e^\mu  \frac{\sin \frac{\pi}{4g} }{ \cos \frac{\pi}{4g} }    \\
    v_j&=e^{j\frac{\pi}{4g}S}v_{4g} \ (j=1, \cdots, 4g-1).
\end{aligned}
\end{align}

\if{

\begin{prop}
There exists a regular $p$-gon with the following points as vertices:
\begin{align}
\begin{aligned}
    v_p&=\frac{\cos \frac{\pi}{q}}{\cos \frac{\pi}{p}}l-\frac{\sin \frac{\pi}{p} }{ \cos \frac{\pi}{p}}+i l  \frac{\sin \frac{\pi}{q} }{ \cos \frac{\pi}{p} } \ \qty(l=\frac{\cos \frac{\pi}{q}}{\sin \frac{\pi}{p}}+\sqrt{  \frac{ \cos^2 \frac{\pi}{q}}{ \sin^2 \frac{\pi}{p}} -1}  )
    \\
    v_n&=e^{n\frac{\pi}{p}S}v_p \ (n=1, \cdots, p-1).
\end{aligned}
\end{align}
And then, $D$ gives the $\{p,q\}$ tiling. In other words, there exists a Fuchsian group generating $D$.
\end{prop}

\begin{prf}
We assume that $D$ is symmetric about the imaginary axis and centered at $z=i$.  \fix{Complete the proof }

\qed
\end{prf}
}\fi

\begin{figure}
    \centering
    \includegraphics[scale=1,bb=  0 0 370 216 ]{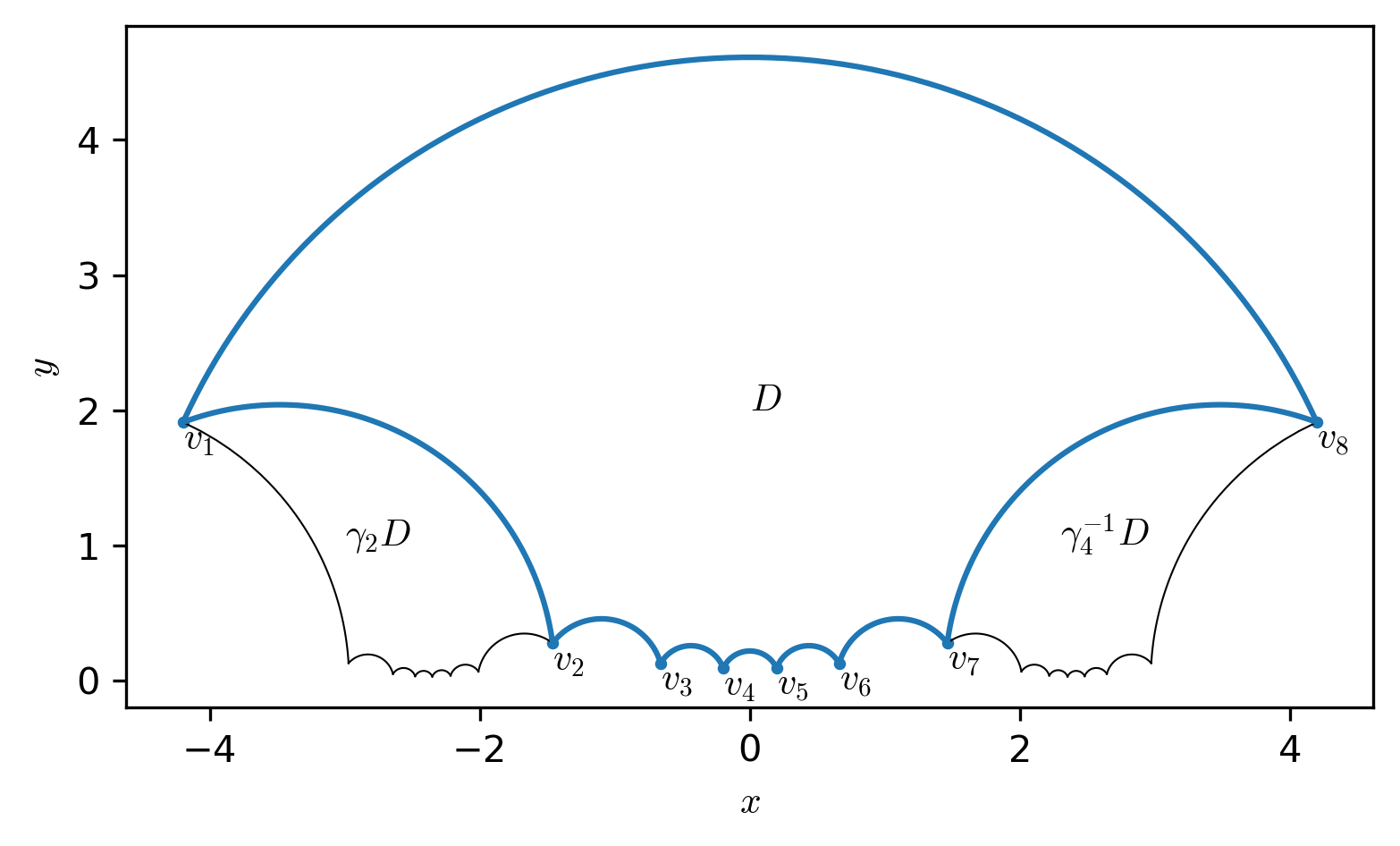}
    \caption{A fundamental domain of $\{8,8\}$ tiling.}
    \label{fig:fundamental domain}
\end{figure}

In particular, a Fuchsian group $\Gamma$ associated with the $\{8,8\}$ tiling is a group generated by the following four elements:
\begin{align}\label{eq:the generators fo Fuchsian grp}
\begin{aligned}
    \gamma_j=&\mqty(\cos( \frac{j-1}{8}\pi) & \sin( \frac{j-1}{8}\pi) \\-\sin( \frac{j-1}{8}\pi) & \cos( \frac{j-1}{8} \pi) )
    \mqty (e^\mu & 0 \\ 0 & e^{-\mu}  )\mqty(\cos( \frac{j-1}{8}\pi) & -\sin( \frac{j-1}{8}\pi) \\\sin( \frac{j-1}{8}\pi) & \cos( \frac{j-1}{8} \pi) )\\
    =&\exp(\mu \qty(U \cos(\frac{j-1}{4}\pi ) -(2T-S) \sin(\frac{j-1}{4}\pi)) ),
\end{aligned}
\end{align}
where $j=1,\cdots,4$ and 
\begin{equation}
    e^\mu=\frac{\cos \frac{\pi}{8}}{\sin \frac{\pi}{8}}+\sqrt{  \frac{ \cos^2 \frac{\pi}{8}}{ \sin^2 \frac{\pi}{8}} -1} =(1+\sqrt{2}) \qty(1+\sqrt{2}\sqrt{ \sqrt{2}-1} ).
\end{equation}
This Fuchsian group $\Gamma$ generates a fundamental domain as Fig. \ref{fig:fundamental domain}.

Below, we consider the Fuchsian group $\Gamma$ that gives $\{4g,4g\}$ tiling, where $g \geq 2$ is an integer. The generators of $\Gamma$ satisfy
\begin{align}
    \gamma_1 \gamma_2^{-1}\cdots \gamma_{2g}^{-1} \gamma_1^{-1} \gamma_2 \cdots \gamma_{2g}=1.
\end{align}
This formula allows us to identify the opposite edges of $D$. Let $C_j\ (j=1,\cdots,4g)$ be edges of $D$ that connects $v_{j-1}$ and $v_{j}$, where $v_{-1}=v_{4g}$. Then each $C_j$ fulfill
\begin{align}
    \gamma_j C_{j+2g}=C_j.
\end{align}
The fundamental domain is regular $4g$-gon, and the Riemann surface of genus $g$ can be constructed by identifying the facing edges. Since the metric $ds^2=\frac{1}{y^2}(dx^2+dy^2)$ is invariant under $\Gamma \subset SL_2(\mathbb{R})$, the metric naturally leads to  the Riemann surface.


The action of $SL_2(\mathbb{R})$ on a function $f$ on $\mathbb{H}$ is determined by
\begin{align}
    (\rho(g)f)(z)=f(gz)~g\in SL_2(\mathbb{R}),z\in \mathbb{H},
\end{align}
where $\rho$ is a representation to a function space.  This $\rho$ reverses the order of products as
\begin{align}
    (\rho(g_1)\rho(g_2) f)(z)=f(g_2 g_1 z)=(\rho(g_2 g_1)f)(z)
\end{align}
for any $g_1,g_2\in SL_2(\mathbb{R})$.

To find the generators of this transformation, we take $g=e^{\theta S},e^{tT},e^{\mu U}$ and differentiate by a parameter, respectively. Then we obtain
\begin{align}
\begin{aligned}
    \eval{\dv{}{\theta} \qty( \rho(e^{\theta S})f) (z) }_{\theta=0} &=  \eval{\dv{}{\theta} f(e^{\theta S} z) }_{\theta=0}=(1+x^2-y^2)\pdv{}{x}f+2xy \pdv{}{y}f \\
    \eval{\dv{}{t} \qty( \rho(e^{t T})f) (z) }_{\theta=0}& =  \eval{\dv{}{t} f(e^{t T} z) }_{\theta=0}=\pdv{}{x}f \\
        \eval{\dv{}{\mu} \qty( \rho(e^{\mu U})f) (z) }_{\theta=0} &=  \eval{\dv{}{\mu} f(e^{\mu U} z) }_{\theta=0}=2x\pdv{}{x}f+2y \pdv{}{y}f .
\end{aligned}
\end{align}
Then we define the generators of the transformation as
\begin{align}\label{eq:the generators of PSL_2}
\begin{aligned}
    \hat{S}&:=(1+x^2-y^2)\pdv{}{x}+2xy\pdv{}{y} \\
    \hat{T}&:=\pdv{}{x} \\
    \hat{U}&:=2x\pdv{}{x}+2y\pdv{}{y}.
\end{aligned}
\end{align}
The commutation relation of $\{\hat{S},\hat{T},\hat{U}\}$ is
\begin{align}\label{eq:commutation relation of trsf}
[\hat{U},\hat{T}]=-2\hat{T},\ [\hat{U},\hat{S}]=-4\hat{T}+2\hat{S},\ [\hat{S},\hat{T}]=-\hat{U}.
\end{align}
This commutation relation differs in sign from the commutation relation \eqref{eq:commutation relation of SL2}. This fact reflects that $\rho$ reverses the order of the products. Thus the transformation corresponding to $g=\exp(X) \in SL_2(\mathbb{R})$ is written by
\begin{align}
    \hat{g}:=\rho(g)=\exp(\hat{X})=\sum_{n=0}^\infty \frac{1}{n!} \hat{X}^n,
\end{align}
where $X$ is an element of $sl_2(\mathbb{R})$.
The transformation corresponding to the generators $\gamma_1\cdots \gamma_4$ of Fuchsian group of the $\{ 8,8\}$ tiling are
\begin{align}\label{eq:transformation by gamma}
\begin{aligned}
    \hat{\gamma_j}:=&\rho(\gamma_j)=\exp(\mu \qty(\hat{U}
    \cos(\frac{j-1}{4}\pi ) -(2\hat{T}-\hat{S}) \sin(\frac{j-1}{4}\pi)) ) \\
    =&\exp( -\frac{j-1}{8}\pi \hat{S}) \exp( \mu \hat{U}) \exp( \frac{j-1}{8}\pi \hat{S}).
\end{aligned}
\end{align}
Then these $\hat{\gamma}_1,\cdots \hat{\gamma}_4$ satisfy
\begin{align}
\begin{aligned}
        \hat{\gamma}_4 \hat{\gamma}_3^{-1}\hat{\gamma}_2 \hat{\gamma}_1^{-1}\hat{\gamma}_4^{-1}\hat{\gamma}_3 \hat{\gamma}_2^{-1} \hat{\gamma}_1=1.
\end{aligned}
\end{align}
Finally we find a group generated by $\{\hat{S},\hat{T},\hat{U}\}$. For arbitrary function $f$, the transformation $e^{\pi \hat{S}}$ act as
\begin{align}
    (e^{\pi \hat{S}} f)(z)=f (e^{\pi S} \cdot z)=f((-1)\cdot z)=f(z)\ (z\in \mathbb{H}).
\end{align}
Therefore $\{\hat{S},\hat{T},\hat{U}\}$ generate
\begin{align}
    PSL_2(\mathbb{R}) \simeq SL_2(\mathbb{R})/\{\pm 1 \},
\end{align}
which means that $SL_2(\mathbb{R})$ is the double-covering of $PSL_2(\mathbb{R})$. 

\subsection{\label{sec:MFG}Magnetic Fuchsian Group}
In this section we give a prescription for constructing the {\it magnetic Fuchsian group}. With respect to the gauge field $A=B\frac{1}{y}dx$ obeying $dA=B \text{vol}$, we work in Landau gauge.

We first consider the $B=0$ case.
\if{The generators of $SL_2(\mathbb{R})$ can be represented as
\begin{align}
\begin{aligned}
        S&=(1+x^2-y^2)\pdv{}{x}+2xy\pdv{}{y}\\
        T&=\pdv{}{x}\\
        U&=2x\pdv{}{x}+2y\pdv{}{y}.
\end{aligned}
\end{align}
They correspond to rotation, translation, and scale transformation, respectively. They respect the commutation relations
\begin{align}
    [U,T]=-2T,\ [U,S]=2S-4T,\ [S,T]=U. 
\end{align} }\fi
By using eq.~\eqref{eq:the generators of PSL_2}, the Hamiltonian can be written as 
\begin{align}
    H=\frac{1}{2m}\qty(\hat{T}(\hat{S}-\hat{T})-\frac{1}{4}\hat{U}^2-\frac{1}{2}\hat{U}). 
\end{align}

Now let us consider the $B\neq0$ case.
\begin{defn}\label{def:magneticSPL}
We put 
\begin{align}
\begin{aligned}\label{eq:generator of magnetic SL2}
    \hat{S}_B&=(1+x^2-y^2)\pdv{}{x}+2xy\pdv{}{y}+2iBy \\
    \hat{T}_B&=\pdv{}{x}=\hat{T} \\
    \hat{U}_B&=2x\pdv{}{x}+2y\pdv{}{y}=\hat{U},
\end{aligned}
\end{align}
and call them generators of the {\it magnetic} $PSL_2(\mathbb{R})$. \end{defn}

They satisfy the commutation relations
\begin{align}\label{eq:commutation relation of trsf in mag}
    [\hat{U}_B,\hat{T}_B]=-2\hat{T}_B,\ [\hat{U}_B,\hat{S}_B]=-4\hat{T}_B+2\hat{S}_B,\ [\hat{S}_B,\hat{T}_B]=-\hat{U}_B
\end{align}
Therefore the Lie algebra $\mathfrak{sl}_2(\mathbb{R})$ of $PSL_2(\mathbb{R})$ and the Lie algebra $\mathfrak{sl}_2(\mathbb{R})[B]$ generated by $\{S_B,T_B,U_B\}$ are isomorphic. We write $PSL_2(\mathbb{R})[B]$ for the {\it magnetic} $PSL_2(\mathbb{R})$, that is the Lie group corresponding to $\mathfrak{sl}_2(\mathbb{R})[B]$.

\begin{lem}
The Hamiltonian \eqref{eq:Hamiltonian} commutes with $\hat{S}_B,\hat{T}_B,\hat{U}_B$.
\end{lem}
One can show this statement by rewriting the Hamiltonian as
\begin{align}
    H=\frac{1}{2m}\qty(\hat{T}_B(\hat{S}_B-\hat{T}_B)-\frac{1}{4}\hat{U}_B^2-\frac{1}{2}\hat{U}_B+B^2). 
\end{align}
For the detail of the derivation of this Hamiltonian, please refer to \cite{COMTET1987185}. This lemma is important to create some conserving quantity with $\{\hat{S}_B,\hat{T}_B,\hat{U}_B\}$. Especially it is important to recall that in a Euclidean lattice, the tight-binding Hamiltonian commutes with the translation operators, hence the Bloch states are their simultaneous eigenstates. To define magnetic Bloch states on a hyperbolic plane under a magnetic field, the property stated in this lemma is important.

Topologically $PSL_2(\mathbb{R})\simeq S^1\times \mathbb{R}^2$, thus $PSL_2(\mathbb{R})$ can be a covering group of $PSL_2(\mathbb{R})[B]$ by choosing an appropriate $B$, which we will discuss later. 
\if{
Since $\hat{S}_B=\hat{S}+2iBy$, $e^{t\hat{S}_B}$ acts on any $ f\in C^\infty (\mathbb{H})$ as
\begin{align}
    (e^{t\hat{S}_B}f)(z)=j(e^{t\hat{S}_B},z) (e^{t\hat{S}}f) (z)=j(e^{t\hat{S}_B},z) f(e^{tS}z) \ (z \in \mathbb{H}),
\end{align}
where $j(e^{t\hat{S}_B},z) \in \mathbb{C}$ is determined by $e^{t\hat{S}_B},z$. }\fi

\begin{prop}
$e^{t\hat{S}_B}$ acts on any $ f\in C^\infty (\mathbb{H})$ as
\begin{align}
    (e^{t \hat{S}_B} f)(z_0)=\exp( \int_0^t 2iB y(t^\prime)dt^\prime) (e^{t\hat{S}}f) (z_0)\ (z_0 \in \mathbb{H}),
\end{align}
where $y(t^\prime)=\Im{e^{t^\prime S} z_0 }$.
Then, $e^{t \hat{S}_B}$ changes not only the position but also the phase.

\end{prop}

\begin{prf}
We define 
\begin{align}
\begin{aligned}
    (\Psi_t f)(z_0)=&\exp( \int_0^t 2iB y(t^\prime)dt^\prime) (e^{t\hat{S}}f) (z_0) \\
    =&\exp( \int_0^t 2iB y(t^\prime)dt^\prime) f (e^{tS}z_0),
\end{aligned}
\end{align}
and then we find
\begin{align}
\begin{aligned}
    (\Psi_{t_1} \Psi_{t_2} f)(z_0)=&\exp(\int_0^
    {t_1}  2iBy_1(t) dt)\qty(\Psi_{t_2} f)(e^{t_1 S}z_0) \\
    =&\exp(\int_0^
    {t_1}  2iBy_1(t) dt) \exp(\int_0^
    {t_2}  2iBy_2(t) dt) f( e^{t_2 S} e^{t_1 S} z_0),
\end{aligned}
\end{align}
where $y_1(t)=\Im{e^{tS}z_0 }\ ,y_2(t)=\Im{ e^{tS}e^{t_1 S} z_0 }$. We define a new curve $y(t)=\Im{e^{tS}z_0 } \ (0\leq t \leq t_1+t_2)$, then
\begin{align}
\begin{aligned}
    y(t)=\Im{e^{tS}z_0}=\left\{ \begin{array}{cc}
        y_1(t) &  (0\leq t \leq t_1) \\
        y_2(t-t_1) & (t_1 \leq t \leq t_2) .
    \end{array}\right.
\end{aligned}
\end{align}
Therefore we obtain
\begin{align}
\begin{aligned}
        (\Psi_{t_1} \Psi_{t_2} f)(z_0)= \exp( \int_0^{t_1+t_2} 2iBy(t) dt)f(e^{(t_1+t_2)S}z_0)=(\Psi_{t_1+t_2} f)(z_0).
\end{aligned}
\end{align}
It means $\Psi_{t_1} \Psi_{t_2}=\Psi_{t_1+t_2}$ and generates a 1-parameter transformation $\{\Psi_t\}_{t\in \mathbb{R}}$. We find $\Psi_t=e^{t\hat{S}_B}$, because it obeys
\begin{align}
        \eval{\dv{t}((\Psi_{t}f)(z) )}_{t=0}=(\hat{S}_B)f(z),~\forall z\in H.
\end{align}

\qed
\end{prf}

Let
\begin{align}\label{eq:the variation of phase}
    j(e^{t\hat{S}_B},z)=\exp( \int_0^t 2iB y(t^\prime)dt^\prime)
\end{align}
be the variation of the phase factor by acting $e^{t\hat{S}_B}$. The variation of the phase has a great deal to do with how many times $PSL_2(\mathbb{R})$ covers $PSL_2(\mathbb{R})[B]$.

The following theorem is one of the main results of the paper.
\begin{thm}
\label{thm:magneticPSL}
If $B=1/q$, $PSL_2(\mathbb{R})$ is a $q$-fold covering of $PSL_2(\mathbb{R})[B]$. 
\end{thm}
\begin{prf}
In $PSL_2(\mathbb{R})$, it satisfy that $e^{\pi\hat{S}}=id$. Therefore by calculating $e^{\pi \hat{S}_B}$, we can see that how many times $PSL_2(\mathbb{R})$ covers $PSL_2(\mathbb{R})[B]$. Then we perform the integral of eq.~\eqref{eq:the variation of phase}. By using eq.~\eqref{eq:differential eq of z(theta)}, we get
\begin{align}
    \int_0^t  2iBy(t^\prime) dt^\prime= \int_0^t 2iB y(t^\prime)\frac{y^\prime}{ y^\prime} dt^\prime=\int_C 2iBy\frac{dy}{2xy}=\int_C iB\frac{dy}{x},
\end{align}
where $y^\prime$ is a derivative of $y(t^\prime)$ and $C=\Set{ e^{t^\prime S} z_0 | 0\leq t^\prime \leq t}$. According to Prop. \ref{prop:orbit of S}, the curve of C is an arc with a center $z=ia$ and radius $b$. Then let
\begin{align}
    x=b\cos(\theta),\ y=b\sin(\theta)+a,
\end{align}
be functions of $\theta$. We define $\theta_i,\theta_f$ as
\begin{align}
    z(\theta_i)=z_0 ,\ z(\theta_f)=e^{tS}z_0.
\end{align}
So, it can be written as an integral by $\theta$, then
\begin{align}
        \int_C iB\frac{dy}{x}=\int_{\theta_i}^{\theta_f}iB d\theta =iB (\theta_f-\theta_i).
\end{align}
Therefore we get
\begin{align}
    j(e^{\pi \hat{S}_B},z)=e^{i2\pi B}.
\end{align}
In particular, this indicates that if $B=\frac{1}{q}$ then $PSL_2(\mathbb{R})[B]\simeq PSL_2(\mathbb{R})/\mathbb{Z}_q$, hence $PSL_2(\mathbb{R})$ is a $q$-fold covering of $PSL_2(\mathbb{R})[1/q]$. 
\qed
\if{
To prove the statement, we derive the general degree of the cover $\pi:SL_2(\mathbb{R})[B] \to SL_2(\mathbb{R})$. For this purpose we define $Exp(\theta S)$ that acts on $\forall f \in C^\infty (\mathbb{H})$ in such a way that 
\begin{align}
    (Exp(\theta S)f)(z)=f\qty (\mqty( \cos\theta & \sin\theta \\ -\sin\theta &\cos\theta )\cdot z)=f(R(\theta)\cdot z). 
\end{align}
We put $S_B=S+V$ and derive the explicit formula \eqref{eq:exp} of $Exp(\theta S_B)$. Here $V=-\frac{iB}{2}\qty(\frac{x^2}{y}-3y-\frac{1}{y})$. We define 
\begin{align}\label{eq:Exp(theta S_B)}
    \Phi_\theta f(z)=\exp( \int^\theta_0 ds V(R(s)\cdot z))f(R(\theta)\cdot z)
\end{align}
and then find 
\begin{align*}
    &(\Phi_{t^\prime} (\Phi_t f))(z)\\
    &=\exp( \int^{t^\prime}_0 ds V(R(s)\cdot z))(\Phi_t f)(R(t^\prime)\cdot z) \\
    &=\exp( \int^{t^\prime}_0 ds V(R(s)\cdot z))\exp(\int_0^t ds V(R(s)\cdot R(t^\prime)\cdot z))f(R(t)\cdot R(t^\prime)\cdot z) \\
    &=\exp(\int^{t^\prime}_0 ds V(R(s)\cdot z)+ \int_{t^\prime}^t ds V(R(s)\cdot z))f(R(t+t^\prime)\cdot z) \\
    &=(\Phi_{t+t^\prime})f(z)
\end{align*}
Therefore they satisfy $\Phi_t \Phi_{t^\prime}=\Phi_{t+t^\prime}$ and generate a 1-parameter transformation $\{\Phi_{t}\}_{t\in \mathbb{R}}$. We find $\Phi_{t}=Exp(tS_B)$, since it obeys 
\begin{align}
    \eval{\dv{t}((\Phi_{t}f)(z) )}_{t=0}=(S_B)f(z),~\forall z\in H.
\end{align}

Note $Exp(\pi S)=id$. To end the prof of the theorem, we calculate $Exp(\pi S_B)$. Using eq.~\eqref{eq:Exp(theta S_B)} we find 
\begin{align}
\begin{aligned}
    Exp(\pi S_B)f(z)&=\exp( \int^\pi_0 ds V(R(s)\cdot z))f(R(\pi)\cdot z) \\
    &=\exp( \int^\pi_0 ds V(R(s)\cdot z))f(z).
\end{aligned}
\end{align}
So the problem is now to find $\exp( \int^\pi_0 ds V(R(s)\cdot z))$. Rewriting $S_B$ as   
\begin{align}
\begin{aligned}
        S_B&=(1+x^2-y^2)\pdv{}{x}+2xy\pdv{}{y}-\frac{iB}{2}\qty(\frac{x^2}{y}-3y-\frac{1}{y}) \\
    &=(1+x^2-y^2)\qty(\pdv{}{x} +\frac{iB}{2y})+2xy \qty(\pdv{}{y}+\frac{iB}{2}\qty(-\frac{x}{y^2}+\frac{2}{x}))
\end{aligned}
\end{align}
we write $V(z)$ as 
\begin{align}
    V(z)=(1+x^2-y^2)\frac{iB}{2y}+2xy\frac{iB}{2}\qty(-\frac{x}{y^2}+\frac{2}{x}).
\end{align}
Therefore we obtain the following formula 
\begin{align}
    \int^\pi_0 ds V(R(s)\cdot z)=\oint_C \qty(\frac{iB}{2y}dx+ \frac{iB}{2}\qty(-\frac{x}{y^2}+\frac{2}{x})dy),
\end{align}
where $C$ is defined by $C=\Set{R(s)\cdot z | s\in [0,\pi] }$ for a given $z$. The direction of $C$ should agrees with that of  $R(s)\cdot z $. One will find that $C$ is a circle whose center is at the imaginary axis:
\begin{align}
    C=\Set{(x,y)\in \mathbb{H} | x^2+(y-\cosh\lambda )^2=\sinh^2 \lambda}
\end{align}
Due to the Stokes theorem, we obtain 
\begin{align}
\begin{aligned}
    &\oint_C \qty(\frac{iB}{2y}dx+ \frac{iB}{2}\qty(-\frac{x}{y^2}+\frac{2}{x})dy)=-iB\int_D dxdy \frac{1}{x^2} \\
    &=-iB \int_{\cosh \lambda -\sinh \lambda}^{\cosh \lambda +\sinh \lambda} dy \int_{-\sqrt{\sinh^2\lambda-(y-\cosh \lambda)^2}}^{\sqrt{\sinh^2\lambda-(y-\cosh \lambda)^2}} dx\frac{1}{x^2} \\
    &=iB\int_{\cosh \lambda -\sinh \lambda}^{\cosh \lambda +\sinh \lambda} dy \frac{2}{\sqrt{\sinh^2\lambda-(y-\cosh \lambda)^2}} \\
    &=iB\int_{-\sinh \lambda}^{\sinh \lambda} dy \frac{2}{\sqrt{\sinh^2\lambda -y^2}} \\
    &=iB2[\sin^{-1}(y)]_{-1}^1
    =iB2\pi,
\end{aligned}
\end{align}
where $D$ is the disc whose boundary is $C$. Therefore we derive the explicit formula 
\begin{equation}
\label{eq:exp}
    Exp(S_B \pi )f=\exp(i2\pi B)f. 
\end{equation} 
In particular, $SL_2(\mathbb{R})$ is a $q$-fold covering of $SL_2(\mathbb{R})[B]$ if $B=\frac{1}{q}$ }\fi 
\end{prf}

For simplicity we consider the Hyperbolic lattice whose tiling is specified by $\{8,8\}$. A Fuchsian group corresponding to the $\{8,8\}$ tiling is defined by eq.~\eqref{eq:the generators fo Fuchsian grp}. Then generators of {\it magnetic Fuchsian group} are defined by
\begin{align}\label{eq:the generators of mag fuchsian grp of 8,8 tiling}
\begin{aligned}
    \hat{\gamma}_j^B=&\exp(\mu \qty(\hat{U}_B
    \cos(\frac{j-1}{4}\pi ) -(2\hat{T}_B-\hat{S}_B) \sin(\frac{j-1}{4}\pi)) )\\
    =&\exp( -\frac{j-1}{8}\pi \hat{S}_B) \exp( \mu \hat{U}_B) \exp( \frac{j-1}{8}\pi \hat{S}_B).
\end{aligned}
\end{align}

\if{Then the Fuchsian group $\Gamma$ is generated by 
\begin{align}
\label{eq:gamma}
    \gamma_j=R\qty((j-1)\frac{\pi}{8})\mqty( \mu & 0 \\ 0& \mu^{-1})R\qty(-(j-1)\frac{\pi}{8}) \ (j=1,2,3,4,),
\end{align}
where $\mu=(1+\sqrt{2})(1+\sqrt{2}\sqrt{\sqrt{2}-1})$. $R(\theta)$ is a rotation matrix
\begin{align}
    R(\theta)=\mqty( \cos\theta & \sin\theta \\ -\sin\theta &\cos\theta )=Exp (\theta S),
\end{align}
where $Exp(\theta S)$ is an exponential map corresponding the the pull-back by the integral curve of the vector field $S$. In other words
\begin{align}
    Exp (\theta S)=\sum_{n=0}^\infty \frac{\theta^n}{n!}S^n
\end{align}
Substituting $S_B$ for $S_B$ into \eqref{eq:gamma}, we write $\gamma^B_j$ for the left hand side of \eqref{eq:gamma}. 
}\fi

\begin{defn}
\label{def:Fuchsian}
We call $\Gamma[B]$ {\it magnetic Fuchsian group} when it is generated by $\hat{\gamma}_j^B$. 
\end{defn}

The same procedure can be used to make $\Gamma[B]$ from general $\Gamma$ of the $\{4g,4g\}$ tiling.

\begin{thm}
\label{thm:Fuchsian}
Let $\Gamma$ be a Fuchsian group of the $\{4g,4g\}$ tiling. The generators $\hat{\gamma}^B_j\ (j=1,\cdots ,2g)$ of magnetic Fuchsian group corresponding to $\Gamma$ satisfy
\begin{align}\label{eq:def eq of mag fuchsian grp}
    \hat{\gamma}_{2g}^B \cdots \hat{\gamma}_2^B (\hat{\gamma}_1^B)^{-1}(\hat{\gamma}_{2g}^B)^{-1}\cdots (\hat{\gamma}_2^B)^{-1} \hat{\gamma}_1^B=e^{i4(g-1)\pi B}.
\end{align}
Here $4(g-1)\pi$ corresponds to the area of the fundamental domain of the $\{4g,4g\}$ tiling. In other words, $\phi=4(g-1)\pi B$ is equal to the magnetic flux through the fundamental domain.
\end{thm}

\begin{prf}
At first, we denote that the left hand side of eq.~ \eqref{eq:def eq of mag fuchsian grp} is a constant. We define a new path that connect $1$ and $\gamma_j$ as
\begin{align}
    \gamma_j(t)=&\mqty(\cos( \frac{j-1}{4g}\pi) & \sin( \frac{j-1}{4g}\pi) \\-\sin( \frac{j-1}{4g}\pi) & \cos( \frac{j-1}{4g} \pi) )
    \mqty (e^t & 0 \\ 0 & e^{-t}  )\mqty(\cos( \frac{j-1}{4g}\pi) & -\sin( \frac{j-1}{4g}\pi) \\\sin( \frac{j-1}{4g}\pi) & \cos( \frac{j-1}{4g} \pi) ),
\end{align}
where $e^t$ satisfies 
\begin{equation}
    1\leq e^t \leq \frac{\cos \frac{\pi}{4g}}{\sin \frac{\pi}{4g}}+\sqrt{  \frac{ \cos^2 \frac{\pi}{4g}}{ \sin^2 \frac{\pi}{4g}} -1} . 
\end{equation}

By using eq.~\eqref{eq:docomposition of SL2}, then we obtain
\begin{align}
    \gamma_1(t) \gamma_2^{-1}(t) \cdots \gamma_{2g}^{-1}(t) \gamma_1^{-1}(t) \gamma_2(t) \cdots \gamma_{2g}(t)=e^{c(t) U} e^{b(t) T} e^{a(t) S},
\end{align}
where $a(0)=b(0)=c(0)=0$. In particular, if $e^t=\frac{\cos \frac{\pi}{4g}}{\sin \frac{\pi}{4g}}+\sqrt{  \frac{ \cos^2 \frac{\pi}{4g}}{ \sin^2 \frac{\pi}{4g}} -1} $, then
\begin{align}
    a(t)=2\pi n, b(t)=c(t)=0\ (n\in \mathbb{Z}).
\end{align}

Similarly for the magnetic Fuchsian group, we get
\begin{align}
    \hat{\gamma}_{2g}^B (t) \cdots \hat{\gamma}_2^B (t) (\hat{\gamma}_1^B (t))^{-1}(\hat{\gamma}_{2g}^B (t))^{-1}\cdots (\hat{\gamma}_2^B (t))^{-1} \hat{\gamma}_1^B (t)=e^{a(t) \hat{S}_B} e^{b(t) \hat{T}_B} e^{c(t) \hat{U}_B} 
\end{align}
Therefore when $e^t= \frac{\cos \frac{\pi}{4g}}{\sin \frac{\pi}{4g}}+\sqrt{  \frac{ \cos^2 \frac{\pi}{4g}}{ \sin^2 \frac{\pi}{4g}} -1} $, then we find 
\begin{align}
        \hat{\gamma}_{2g}^B \cdots \hat{\gamma}_2^B (\hat{\gamma}_1^B)^{-1}(\hat{\gamma}_{2g}^B)^{-1}\cdots (\hat{\gamma}_2^B)^{-1} \hat{\gamma}_1^B=e^{2\pi n \hat{S}_B}=e^{i 4\pi n B}. 
\end{align}
Here $e^{4\pi n B}$ is a constant. 

We show $n=g-1$ in rest of the proof. Then we transform the left-hand side of eq.~\eqref{eq:def eq of mag fuchsian grp} as in 
\begin{align}
\begin{aligned}
        &\hat{\gamma}_{2g}^B \cdots \hat{\gamma}_2^B (\hat{\gamma}_1^B)^{-1}(\hat{\gamma}_{2g}^B)^{-1}\cdots (\hat{\gamma}_2^B)^{-1} \hat{\gamma}_1^B\\
    =&e^{-\frac{2g-1}{4g} \pi \hat{S}_B} \underbrace{(e^{\mu \hat{U}_B}e^{\frac{1}{4g} \pi \hat{S}_B}e^{-\mu \hat{U}_B}e^{\frac{1}{4g} \pi \hat{S}_B}) \cdots (e^{\mu \hat{U}_B}e^{\frac{1}{4g} \pi \hat{S}_B}e^{-\mu \hat{U}_B}e^{\frac{1}{4g} \pi \hat{S}_B}) }_{g\text{ terms}}e^{- \frac{1}{4g} \pi \hat{S}_B} \nonumber \\
    &\times  e^{-\frac{2g-1}{4g} \pi \hat{S}_B} \underbrace{(e^{-\mu \hat{U}_B}e^{\frac{1}{4g} \pi \hat{S}_B}e^{\mu \hat{U}_B}e^{\frac{1}{4g} \pi \hat{S}_B}) \cdots (e^{-\mu \hat{U}_B}e^{\frac{1}{4g} \pi \hat{S}_B}e^{\mu \hat{U}_B}e^{\frac{1}{4g} \pi \hat{S}_B}) }_{g\text{ terms}}e^{- \frac{1}{4g} \pi \hat{S}_B}
\end{aligned}
\end{align}

Let
\begin{align}
    j(e^{\frac{\pi}{4g} \hat{S}_B},v_j)=e^{iB \theta_{j+1}}
\end{align}
be a the change in phase with respect to the vertex $v_1,\cdots v_{4g}$ defined by eq.~\eqref{eq:vertex of D} of $D$ by acting $e^{\frac{\pi}{4g} \hat{S}_B}$. Thus $ \hat{\gamma}_{2g}^B \cdots \hat{\gamma}_2^B (\hat{\gamma}_1^B)^{-1}(\hat{\gamma}_{2g}^B)^{-1}\cdots (\hat{\gamma}_2^B)^{-1} \hat{\gamma}_1^B$ acts on any $ f\in C^\infty (\mathbb{H})$ at $z=v_{4g}$ as
\begin{align}
\begin{aligned}
    &\hat{\gamma}_{2g}^B \cdots \hat{\gamma}_2^B (\hat{\gamma}_1^B)^{-1}(\hat{\gamma}_{2g}^B)^{-1}\cdots (\hat{\gamma}_2^B)^{-1} \hat{\gamma}_1^B f(v_{4g})\\
    =&\exp(iB( (2g-1)(\theta_1+\theta_{2g+1})-(\theta_2+\theta_3+\cdots+\theta_{2g})-(\theta_{2g+2}+\cdots+\theta_{4g} ) )) f(v_{4g}).
\end{aligned}
\end{align}
Here $\theta_2+\theta_3+\cdots+\theta_{2g}$ is equal to $\theta_{2g+2}+\cdots+\theta_{4g}$. So we simply write $\theta=\theta_2+\theta_3+\cdots+\theta_{2g}$. By using $\theta_1 +\theta_{2g+1}+2\theta=2\pi$, we obtatin
\begin{align}
\begin{aligned}
    &\hat{\gamma}_{2g}^B \cdots \hat{\gamma}_2^B (\hat{\gamma}_1^B)^{-1}(\hat{\gamma}_{2g}^B)^{-1}\cdots (\hat{\gamma}_2^B)^{-1} \hat{\gamma}_1^B f (v_{4g})\\
    =&\exp(iB( (2g-1)(\theta_1+\theta_{2g+1})-2\theta ))f(v_{4g})\\
    =&\exp(iB( (4g-2)\pi -4g \theta ) )f(v_{4g}),
\end{aligned}
\end{align}
where $\theta$ is defined by $j(e^{\frac{2g-1}{4g} \hat{S}_B},v_1)=e^{iB\theta}$. From eq.~\eqref{eq:the variation of phase}, we get
\begin{align}
     j(e^{\frac{2g-1}{4g}\pi \hat{S}_B },p_1 )=\exp(2iB \int_0^{\frac{2g-1}{4g}\pi } y(t) dt ), 
\end{align}
where $y(t)=\Im{e^{t \hat{S}} p_1 }=\Im \frac{\cos t p_1 +\sin t }{-\sin t p_1 +\cos t }$. Therefore this integral is
\begin{align}
\begin{aligned}
\int_0^{\frac{2g-1}{4g}\pi } y(t) dt=&\int_0^{\frac{2g-1}{4g}\pi }\Im \frac{\cos t p_1 +\sin t }{-\sin t p_1 +\cos t } dt\\
=&-\Im{ \log ( -\sin (\frac{2g-1}{4g}\pi) p_1 + \cos(\frac{2g-1}{4g}\pi)) } \\
=&-\tan[-1]( \frac{y_1}{x_1-\tan(\frac{1}{4g} \pi)} ),
\end{aligned}
\end{align}
where $v_1=x_1+iy_1$. From eq.~\eqref{eq:vertex of D}, the coordinate of $v_1$ is
\begin{align}
\begin{aligned}
    x_1&=-e^{\mu}+\tan( \frac{1}{4g}\pi) \\
    y_1&=e^{\mu}\tan( \frac{1}{4g}\pi) \\
    e^\mu&=\frac{1}{\tan{ \frac{\pi}{4g}}}+ \sqrt{\frac{1}{\tan^2{ \frac{\pi}{4g}}} -1 } .
\end{aligned}
\end{align}
Then we obtain
\begin{align}
\begin{aligned}
j(e^{\frac{2g-1}{4g}\pi \hat{S}_B },p_1 )=&\exp(-i2B \tan[-1](\frac{y_1}{x_1-\tan(\frac{1}{4g} \pi)} )) \\
=&\exp(iB\frac{\pi}{2g}),\ \theta=\frac{\pi}{2g}.
\end{aligned}
\end{align}

Since $\hat{\gamma}_{2g}^B \cdots \hat{\gamma}_2^B (\hat{\gamma}_1^B)^{-1}(\hat{\gamma}_{2g}^B)^{-1}\cdots (\hat{\gamma}_2^B)^{-1} \hat{\gamma}_1^B$ is a constant, then we prove
\begin{align}
\begin{aligned}
    \hat{\gamma}_{2g}^B \cdots \hat{\gamma}_2^B (\hat{\gamma}_1^B)^{-1}(\hat{\gamma}_{2g}^B)^{-1}\cdots (\hat{\gamma}_2^B)^{-1} \hat{\gamma}_1^B =&\exp(iB((4g-2)\pi-2\pi ) ) \\
    =&\exp(i4(g-1)\pi B ).
\end{aligned}
\end{align}
\qed
\end{prf}
This is the Gauss-Bonnet theorem on a hyperbolic surface under a magnetic field. (Please remember the relation between the Aharonov-Bohm phase and topology, where the magnetic field $B$ corresponds to the curvature.) In the physics literature, it corresponds to a magnetic flux characterized by a Wilson line/Berry phase. 

\if{
It is not hard to show that 
\begin{align}
\label{eq:MFG}
    \gamma_1^B(\gamma_2^B)^{-1}\gamma_3^B(\gamma_4^B)^{-1}(\gamma_1^B)^{-1}\gamma_2^B(\gamma_3^B)^{-1}\gamma_4^B=\exp(i4\pi B)
\end{align}
It is important to note that the area of a unit lattice in the   $\{8,8\}$tiling is $4\pi$, so the right hand side of the eq.~\eqref{eq:MFG} explains the flux perpendicular to the unit lattice. As a consequence, we succeed in finding the completely parallel formalism of the conventional ones in the Euclidean lattice: $T_x^B T_y^B=e^{i\phi}T_y^B T_x^B$.  }\fi

\section{\label{sec:butt}Hyperbolic Bloch States under Magnetic Field}
A generic energy spectrum of electrons can be obtained by magnetic Bloch states. In this section, we explain how to construct magnetic Bloch states on a hyperbolic lattice under a magnetic field. Since we are interested in applications to Riemann surface, we work with $\{4g,4g\}$-tiling. This section is devoted for a technical summary that we used in \cite{2021arXiv210413314I} to derive the Hamiltonian to address the Hofstadter butterfly on a hyperbolic lattice. For the detail of motivations and physics background, please refer to the original article~\cite{2021arXiv210413314I}. Moreover the Hofstadter problem on a hyperbolic lattice is studied in a slightly different set up by \cite{2022PhRvL.128p6402S} Discretization of the Hamiltonian on a hyperbolic lattice is also addressed in~\cite{2020arXiv200805489M}, for example. 

Using the generators \eqref{eq:generator of magnetic SL2} of $SL_2(\mathbb{R})[B]$, we write 
\begin{equation}
    \hat{J}_1=\frac{1}{2}(2\hat{T}_B-\hat{S}_B),~\hat{J}_2=\frac{i}{2}\hat{S}_B,~\hat{J}_3=\frac{1}{2}\hat{U}_B, 
\end{equation}
which satisfies the commutation relation 
\begin{equation}
    [\hat{J}_i,\hat{J}_j]=i\epsilon_{ijk} \hat{J}_k. 
\end{equation}
Using them, we can write the Hamiltonian as 
\begin{equation}
\label{eq:Ham}
    H=\frac{1}{2m}\left(-\sum_{i=1}^3 \hat{J}^2_i+B^2\right). 
\end{equation}
Then, by defining the lattice Hamiltonian $H_\text{lattice}$ as 
\begin{equation}
    H_\text{lattice}=-\sum_{i=1}^3\frac{1}{a^2_i}\left(\exp(a_i \hat{J}_i)+\exp(-a_i\hat{J}_i)\right). 
\end{equation}
When 0-th power of the exponential series is neglected, we find it converges into $H$~\eqref{eq:Ham} (up to additive and multiplicative constants) in the limit of $a_i\to 0$.  

Let us tune the lattice spacing $a_i$. For this purpose, we use the formula 
\begin{equation}
    e^{-\theta \hat{S}_B} e^{\mu \hat{U}_B} e^{\theta \hat{S}_B}=\exp(\mu \qty(\cos(2\theta)\hat{U}_B- \sin(2\theta)(2\hat{T}_B-\hat{S}_B)))
\end{equation}
and find 
\begin{align}
\begin{aligned}
    \exp(2\mu \hat{J}_1)&=\exp( \mu (2\hat{T}_B-\hat{S}_B)) =e^{-\frac{\pi}{8}\hat{S}_B}e^{-\mu \hat{U}_B}e^{\frac{\pi}{4} \hat{S}_B} =(\hat{\gamma}_3^B)^{-1} \\
    \exp(-i\frac{\pi}{4}\hat{J}_2)&=\exp(\frac{\pi}{8} \hat{S}_B)\\
    \exp(2\mu \hat{J}_3)&=\hat{\gamma}_1^B,
\end{aligned}    
\end{align}
where $e^\mu=(1+\sqrt{2}) \qty(1+\sqrt{2}\sqrt{ \sqrt{2}-1} ) $.
The Hamiltonian transformed in this way respects the original $\{8,8\}$ tiling:
\begin{align}\label{eq:H lattice}
    H_{\text{lattice}}=-\frac{1}{4\mu^2}\qty(\hat{\gamma}_1^B+(\hat{\gamma}_1^B)^{-1}+\hat{\gamma}_3^B+(\hat{\gamma}_3^B)^{-1})+\frac{4^2}{\pi^2}\qty(\exp(\frac{\pi}{8} \hat{S}^B) +\exp(-\frac{\pi}{8}\hat{S}^B ) )
\end{align}
\if{Alternatively one can use the Hamiltonian that approximates the continuum theory 
\begin{align}\label{eq:isotropic H lattice}
\begin{aligned}
H_\text{lattice}=-\frac{1}{8\mu^2}\sum_{i=1}^4\left(\hat{\gamma}^B_i+\left(\hat{\gamma}^B_i\right)^{-1}\right)+\frac{4^2}{\pi^2}\left(e^{\frac{\pi}{8}\hat{S}_B}+e^{-\frac{\pi}{8} \hat{S}_B}\right),
\end{aligned}    
\end{align}
whose first term has a resemble form of the traditional tight-binding Hamiltonian on a torus $(g=1)$. However, due to the negative curvature of the hyperbolic space, the energy spectrum has a different structure in general. In what follows, we solve the characteristic equation of those Hamiltonians and plot the spectrum as a function of magnetic flux. }\fi

\begin{thm}
Let $\phi=4(g-1)\pi B= 4\pi B$ be a magnetic flux through a regular $8$-gon, where $B= p /2q$ with co-primes $p,q$. Then there exists functions $\psi_{n,k}\ (n=0,\cdots q-1, k=(k_1,k_2,k_3,k_4))$ satisfying
\begin{align}\label{eq:boundary condition of psi_nk}
\begin{aligned}
        \hat{\gamma}_1^B \psi_{n,k}(z)&=j(\hat{\gamma}_1^B,z) \psi_{n,k}({\gamma}_1 z) =e^{ik_1}\psi_{n+1,k}(z) \\
        (\hat{\gamma}_1^B)^{q} \psi_{n,k}(z)&=e^{ik_1 q}\psi_{n+q,k}(z)=e^{ik_1 q}\psi_{n,k}(z) \\
    \hat{\gamma}_2^B \psi_{n,k}(z)&=j(\hat{\gamma}^B_2,z)\psi_{n,k}({\gamma}_2 z)=e^{ik_2-in\phi}\psi_{n,k}(z)\\
        \hat{\gamma}_a^B \psi_{n,k}(z)&=j(\hat{\gamma}^B_a,z)\psi_{n,k}({\gamma}_a z)=e^{ik_j}\psi_{n,k}(z)\ (a=3,4),
\end{aligned}
\end{align}
where $j(\gamma,z)$ is a $U(1)$-representation of $\pi_1(\Sigma_g)$ with $\gamma$ acting on $z$. 
\end{thm}

\begin{prf}
According to Prop. \ref{prop:def of fundamental domain}, $\Gamma$ gives the $\{8,8\}$ tiling as $\mathbb{H}=\bigcup_{\gamma \in \Gamma} \gamma D$. Thus, it suffices to show that there exist functions $\psi_{n,k}$ in $D$ satisfying eq.~\eqref{eq:boundary condition of psi_nk} and are consistent with eq.~\eqref{eq:def eq of mag fuchsian grp}. Let $C_j\ (j=1,\cdots ,8)$ are the edges of $D$, then
\begin{align}
    \gamma_j C_{j+4}=C_j~(j=1,2,3,4).
\end{align}
Thus, we find functions $\psi_{n,k}$ in $D$ satisfying that
\begin{align}
\begin{aligned}
    j(\hat{\gamma}_1^B,z) \psi_{n,k}(C_1) &=e^{ik_1}\psi_{n+1,k}(C_5) \\
    j(\hat{\gamma}^B_2,z)\psi_{n,k}(C_2)&=e^{ik_2-in\phi}\psi_{n,k}(C_6) \\
    j(\hat{\gamma}^B_a,z)\psi_{n,k}(C_a)&=e^{ik_j}\psi_{n,k}(C_{a+4})\ (a=3,4).
\end{aligned}
\end{align}

It follows that $\psi_{n,k}$ is consistent with \eqref{eq:def eq of mag fuchsian grp}:
\begin{align}
\begin{aligned}
     &\hat{\gamma}_4^B (\hat{\gamma}_3^B)^{-1}\hat{\gamma}_2^B (\hat{\gamma}_1^B)^{-1}(\hat{\gamma}_4^B)^{-1}\hat{\gamma}_3^B (\hat{\gamma}_2^B)^{-1} \hat{\gamma}_1^B \psi_{n,k} \\
     &=    \hat{\gamma}_4^B (\hat{\gamma}_3^B)^{-1}\hat{\gamma}_2^B (\hat{\gamma}_1^B)^{-1}(\hat{\gamma}_4^B)^{-1}\hat{\gamma}_3^B (\hat{\gamma}_2^B)^{-1}e^{ik_1}\psi_{n+1,k}  \\
     &= \hat{\gamma}_4^B (\hat{\gamma}_3^B)^{-1}\hat{\gamma}_2^B (\hat{\gamma}_1^B)^{-1}e^{ik_1-ik_2+ik_3-ik_4+i(n+1)\phi}\psi_{n+1,k} \\
     &=\hat{\gamma}_4^B (\hat{\gamma}_3^B)^{-1}\hat{\gamma}_2^B e^{-ik_2+ik_3-ik_4+i(n+1)\phi}\psi_{n,k} \\
     &=e^{-in\phi +i(n+1)\phi}\psi_{n,k } =e^{i\phi} \psi_{n,k}.
\end{aligned}
\end{align}
\qed
\end{prf}

\if{
We can write as $\psi_{n,k}(z)=\psi_k(z) u_n(z)$, where $\psi_{k}$ is a hyperbolic Bloch state \cite{2020arXiv200805489M}, which $\psi_{k}$ satisfies 
\begin{align}
    \hat{\gamma_j}\psi_{k}(z)=\psi_k(\gamma_j z)=e^{k_j} \psi_k( z)\ (j=1,2,3,4)
\end{align}
for generators $\gamma_j$ of a Fuchsian group $\Gamma$. We assume that $u_n(z)$ fulfills
\begin{align}
\begin{aligned}
        \hat{\gamma}_1^B u_{n}(z)&=j(\hat{\gamma}_1^B,z) u_{n}({\gamma}_1 z) =u_{n+1}(z) \\
        (\hat{\gamma}_1^B)^{q} u_n(z)&=u_{n+q}(z)=u_n(z) \\
    \hat{\gamma}_2^B u_n (z)&=j(\hat{\gamma}^B_2,z)u_n({\gamma}_2 z)=e^{-in\phi}u_n(z)\\
        \hat{\gamma}_a^B u_n(z)&=j(\hat{\gamma}^B_a,z)u_n({\gamma}_a z)=u_n(z)\ (a=3,4).
\end{aligned}
\end{align}
This condition is consistent with \eqref{eq:boundary condition of psi_nk}. }\fi

In order to diagonalize the Hamiltonian \eqref{eq:H_latt}, we compute with the basis 
\begin{align}
\begin{aligned}
    \left(\psi_{0,k},e^{\frac{\pi}{8}\hat{S}_B}\psi_{0,k},\cdots, e^{\frac{7\pi}{8}\hat{S}_B }\psi_{0,k}, \psi_{1,k},e^{\frac{\pi}{8}\hat{S}_B}\psi_{1,k},\cdots, e^{\frac{7\pi}{8}\hat{S}_B}\psi_{1,k},\cdots,\right.  \\
    \left. \psi_{q-1,k},e^{\frac{\pi}{8}\hat{S}_B}\psi_{q-1,k},\cdots, e^{\frac{7\pi}{8}\hat{S}_B}\psi_{q-1,k}  \right)
\end{aligned}
\end{align}
It is important to recall that $R(\theta)$ acts on $\psi(x)=\exp(inx),~n\in\mathbb{Z},$ in such a way that $R(\theta)\exp(inx)=\exp(in(2\theta+x))$. Then the Hamiltonian can be represented as 
\begin{align}\label{eq:H_latt}
    H_\text{lattice}=\left( \begin{array}{cccccc}
A_0 & B^\dagger &  &  &  & B \\
B & A_1 & B^\dagger &  &  &  \\
 & B & A_2 &  &  &  \\
 &  &  & \ddots &  &  \\
 &  &  &  & A_{q-2} & B^\dagger \\
B^\dagger &  &  &  & B & A_{q-1} \\
    \end{array}
    \right),
\end{align}
where $A_n,B~(n=1,\cdots,q-1)$ are the following 8 by 8 matrices 
\begin{align}
    A_n=&-\frac{2}{8\mu^2} \qty(\cos(k_2-n\phi) + \cos(k_3)+ \cos(k_4))I_{8\times 8} \nonumber \\
    &+\frac{4^2}{\pi^2}\left( \begin{array}{cccccccc}
0 & 1 &  &  &  &  &  & e^{i2\pi B} \\
1 & 0 & 1 &  &  &  &  &  \\
 & 1 & 0 & 1 &  &  &  &  \\
 &  & 1 & 0 & 1 &  &  &  \\
 &  &  & 1 & 0 & 1 &  &  \\
 &  &  &  & 1 & 0 & 1 &  \\
 &  &  &  &  & 1 & 0 & 1 \\
e^{-i2\pi B} &  &  &  &  &  & 1 & 0 \\
    \end{array}
    \right) \\
    B=&-\frac{e^{ik_1}}{8\mu^2} I_{8\times 8}
\end{align}

Since $H_\text{lattice}$ and $e^{\frac{\pi}{8}\hat{S}_B}+e^{-\frac{\pi}{8}\hat{S}_B}$ are commutative, there exists simultaneous eigenstates. $e^{\frac{\pi}{8}\hat{S}_B}$ satisfies
\begin{align}
    (e^{\frac{\pi}{8}\hat{S}_B})^8=e^{2\pi B},
\end{align}
so $e^{\frac{\pi}{8}\hat{S}_B}$ has eight eigenvalues $\exp(\frac{i2\pi B}{8} +\frac{i2\pi m }{8})\ (m=0,\cdots , 7)$. Then we obtain the eigenvalues of $e^{\frac{\pi}{8}\hat{S}_B}+e^{-\frac{\pi}{8}\hat{S}_B}$ as
\begin{align}
    2\cos(\frac{\pi B}{4}+\frac{m\pi}{4} )\ (m=0,\cdots , 7).
\end{align}

Therefore the explicit representation of the Hamiltonian \eqref{eq:H_latt} is 
\begin{align}
    H_{\text{lattice}}=&-\frac{1}{8\mu^2} \left( \begin{array}{ccccc}
2\cos(k_2) & e^{-ik_1} &  &  & e^{ik_1} \\
e^{ik_1} & 2\cos(k_2-\phi) &  &  &  \\
 &  & \ddots &  &  \\
 &  &  & 2\cos(k_2-(q-2)\phi) & e^{-ik_1} \\
e^{-ik_1} &  &  & e^{ik_1} & 2\cos(k_2-(q-1)\phi)
    \end{array}
    \right) \nonumber\\
    &-\frac{2}{8\mu^2}(\cos(k_3)+\cos(k_4))+2\cos(\frac{\pi B}{4}+\frac{m\pi}{4} ) .
\end{align}

In addition the Hamiltonian $\eqref{eq:H lattice}$ can be represented as \eqref{eq:H_latt}, where $A_n ,B (n=0,\cdots q-1)$ are determined by
\begin{align}
    A_n=&-\frac{2}{4\mu^2} \mqty(\cos(k_3) &  &  &  &  \\
 & \cos(k_2-n\phi)+\cos(k_4) &  &  &  \\
 &  & \cos(k_3) &  &  \\
 &  &  & \cos(k_2-n\phi)+\cos(k_4) &  \\
 &  &  &  & \ddots ) \nonumber \\
 &+\frac{4^2}{\pi^2}\left( \begin{array}{cccccccc}
0 & 1 &  &  &  &  &  & e^{i2\pi B} \\
1 & 0 & 1 &  &  &  &  &  \\
 & 1 & 0 & 1 &  &  &  &  \\
 &  & 1 & 0 & 1 &  &  &  \\
 &  &  & 1 & 0 & 1 &  &  \\
 &  &  &  & 1 & 0 & 1 &  \\
 &  &  &  &  & 1 & 0 & 1 \\
e^{-i2\pi B} &  &  &  &  &  & 1 & 0 \\
    \end{array}
    \right) \\
B&=-\frac{e^{ik_1}}{4\mu^2}\mqty(1 &  &  &  &  &  &  &  \\
 & 0 &  &  &  &  &  &  \\
 &  & 1 &  &  &  &  &  \\
 &  &  & 0 &  &  &  &  \\
 &  &  &  & 1 &  &  &  \\
 &  &  &  &  & 0 &  &  \\
 &  &  &  &  &  & 1 &  \\
 &  &  &  &  &  &  & 0).
\end{align}

\if{
We consider the tight binding Hamiltonian 
\begin{equation}
    H=-t\sum_{mn}(c^\dagger_ne^{i\theta_{mn}}c_n+h.c), 
\end{equation}
where $\theta_{mn}=\int_{r_m}^{r_n} Adr$ and $m$ is a label for a site $m=(x_m,y_m)$. Let $\gamma_i\in\Gamma$ be a Fuchsian group generator. Then the Schr\"{o}dinger equation is 
\begin{equation}
    -t\sum_{i}\left(e^{i\theta_{\gamma_ix,\gamma_iy}}\psi(\gamma_ix,\gamma_iy)+e^{-i\theta_{\gamma^{-1}_ix,\gamma^{-1}_iy}}\psi(\gamma^{-1}_ix,\gamma^{-1}_iy\right)=E\psi(x,y)
\end{equation}

The local electric current $J_{nm}$ from site $m$ to site $n$ is calculated by
\begin{equation}
\label{eq:current}
J_{nm}=i \frac{(-e)t}{\hbar}(e^{i \theta _{mn}}\psi^*_n\psi_m-{\rm c.c.}).
\end{equation}

We consider the case $\Phi=\frac{p}{q}$. The Hamiltonian can be written with $2g$ generators $T_i,(i=1,\cdots, 2g)$ of quantum group $U_q(sl_{2g})$, that satisfy 
\begin{align}
    [T_{2i},T_{2i+1}]\neq 0&, [T_{2i},T_{2j}]=[T_{2i+1},T_{2j+1}]=0\\ T_1T_2\cdots T_{2g-1}T_{2g}&=e^{ig\Phi}T_{2g}T_{2g-1}\cdots T_2T_1
\end{align}

\begin{equation}
    T_i=\frac{(e_i-e^{-1}_i)k_{i}}{q-q^{-1}}
\end{equation}

\begin{equation}
    H=-t\sum_{i=1}^{2g}T_i+T^\dagger_i.
\end{equation}

Recall that BZ is homeomorphic to $T^{2g}$ and the cuurent runs to the $x$-direction. For $\ket{k}=\ket{k_{x1},k_{y_1},\cdots,k_{xg},k_{yg}}$,
\begin{equation}
    H\ket{k}=E(k)\ket{k},~\text{with}~T_{2i}\ket{k}=e^{iqk_{xi}a}\ket{k}~T^q_{2i+1}\ket{k}=e^{ik_{yi}a}\ket{k} 
\end{equation}

We can construct similar operators that commute with $H$. Each $\tilde{T}_{2i}$ and $\tilde{T}_{2i+1}$ must satisfy 
\begin{equation}
\tilde{T}_{2i}\tilde{T}_{2i+1}=e^{ie\phi}\tilde{T}_{2i+1}\tilde{T}_{2i},\phi=\frac{p}{q} 
\end{equation}

\begin{equation}
    \tilde{T}_{2i+1}\ket{k}\sim \ket{k_{x1},k_{y_1},k_{xi},k_{yi}+2\pi\frac{p}{qa},\cdots,k_{xg},k_{yg}}
\end{equation}

The magnetic BZ is 
\begin{equation}
    -\frac{\pi}{a}\le k_{xi}\le \frac{\pi}{a},~-\frac{\pi}{qa}\le k_{yi}\le \frac{\pi}{qa}
\end{equation}

\begin{align}
    H(k)&=\bigotimes_{i} H_i\\
    H_i&=
\begin{pmatrix}
2\cos(k_{xi}-2\pi\phi)&1&0&\cdots& e^{-iqk_{yi}}\\
1&2\cos(k_{xi}-4\pi\phi)&1&0&0\\
0&1&2\cos(k_{xi}-6\pi\phi)&1&\vdots\\
0&\cdots&\ddots&\ddots&\vdots\\
e^{iqk_{yi}}&0&\cdots&1&2\cos(k_{xi}-2\pi q\phi)
\end{pmatrix}
\end{align}

\begin{figure}
\begin{minipage}{0.32\hsize}
    \centering
    \includegraphics[width=\hsize]{H_butt_g1_97.png}
\end{minipage}
\begin{minipage}{0.32\hsize}
    \centering
    \includegraphics[width=\hsize]{H_butt_g2_53.png}
\end{minipage}
\begin{minipage}{0.32\hsize}
    \centering
    \includegraphics[width=\hsize]{H_butt_g3_23.png}
\end{minipage}
    \caption{[Left]  $g=1$ [Middle] $g=2$ [Right] $g=3$}
    \label{fig:my_label}
\end{figure}
}\fi

\section{Automorphic Forms}
In this section, we give a remark on a connection between the magnetic Bloch states $\psi_{n,k}$~\eqref{eq:boundary condition of psi_nk} and automorphic forms~\cite{venkov1983spectral,10.2307/2372521}. This will be important for investigating further mathematical aspects of the hyperbolic band theory. Euclidean band theoretical cases are studied in~\cite{IKEDA2018136,2018arXiv181211879I,doi:10.1063/1.4998635}. A similar discussion without magnetic field is given in \cite{doi:10.1073/pnas.2116869119}. 

One of the most important connection between automorphic forms and other subjects of mathematics is the Langlands program, which is proposed by a Canadian mathematician, Robert Langlands~\cite{10.1007/BFb0079065}. This conjecture suggests a correspondence between zeta functions of automorphic forms and elliptic curves. Here a physics interpretation of a zeta function is given by partition function, therefore the correspondence implies the duality of two different partition functions. For example, the most fundamental interpretation of this duality is electric-magnetic duality~\cite{2007CNTP....1....1K}, which is sometimes called $S$-duality. In terms of condensed matter physics, this means the duality between an electric charge $e$ and a magnetic monopole with charge $1/e$. In the literature of statistical mechanics, this is the same as Kramers–Wannier duality, which also asserts the correspondence between two partition functions at high temperature $T$ and low temperature $1/T$. For more details, please see \cite{2018arXiv181211879I}. In terms of hyperbolic band theory, those are studied in   \cite{PhysRevE.106.034114,kienzle2022hyperbolic} from a viewpoint of mathematical physics. Moreover, it will be an interesting open problem to consider Kramers–Wannier duality on a hyperbolic surface~\cite{2022arXiv221007227P}.  

We define an automorphic factor $j$ by
\begin{align}
    j(g,z)=\frac{1}{cz+d},~\ \qty(g=\mqty(a & b \\ c &d) \in SL_2 (\mathbb{R}) , z\in \mathbb{H}).
\end{align}
Then $j$ satisfies
\begin{align}
    j(g_1 g_2,z)=j(g_1,g_2 z)j(g_2,z)
\end{align}
for any $g_1 ,g_2 \in SL_2(\mathbb{R}),z \in \mathbb{H}$.

\begin{defn}
If a function $f$ on $\mathbb{H}$ satisfies three following conditions,
\begin{enumerate}
    \item $f$ is a regular function on $\mathbb{H}$,
    \item $f(\gamma z) j(\gamma,z)^m =f(z) \ (\forall \gamma \in \Gamma , \forall z \in \mathbb{H})$,
    \item $f$ is finite at all cusp points,
\end{enumerate}
then $f$ is called an automorphic form of weight $m \in \mathbb{Z}$ for a given Fuchsian group $\Gamma$. 
\end{defn}

Here we consider new generators as
\begin{align}
    \check{S}&=(1+x^2-y^2)\pdv{}{x}+2xy \pdv{}{y}+2B(x+iy) \\
    \check{T} &=\pdv{}{x} \\
    \check{U} &=2x\pdv{}{x}+2y\pdv{}{y}+2B.
\end{align}
The commutation relation of $\{ \check{S} ,\check{T},\check{U}\}$ is
\begin{align}
    [\check{U},\check{T}]=-2\check{T},\ [\check{U},\check{S}]=-4\check{T}+2\check{S},\ [\check{S},\check{T}]=-\check{U},
\end{align}
so a Lie algebra generated by $\{ \check{S} ,\check{T},\check{U}\}$ is isotropic to $\mathfrak{sl}_2(\mathbb{R})$. And then we define a new Hamiltonian $H$ that is commutative with $\{ \check{S} ,\check{T},\check{U}\}$ as
\begin{align}
\begin{aligned}
    H^\prime=&\frac{y^2}{2m} \qty( \qty( -i\pdv{}{x}-\frac{B}{y} )^2 +\qty( -i\pdv{}{y}-i\frac{B}{y} )^2 ) \\
    =&\frac{1}{2m} \qty(\check{T}(\check{S}-\check{T})-\frac{1}{4}\hat{U}^{\prime2}-\frac{1}{2}\check{U}+B^2). 
\end{aligned}
\end{align}

In the same discussion as above, the transformation generated by $ \check{S} $ to a function $f$ on $\mathbb{H}$ is
\begin{align}
    e^{t \check{S} }f(z_0)=\exp( 2B\int_0^t dt^\prime (x(t^\prime)+iy(t^\prime))  )f(e^{tS}z_0),
\end{align}
where $z(t^\prime)=x(t^\prime)+iy(t^\prime)= e^{t^\prime} z_0$. If $2B \in \mathbb{Z}$, then
\begin{align}
    \exp( 2B\int_0^t dt^\prime (x(t^\prime)+iy(t^\prime))  )=&\exp( -2B \log( -\sin(t) z+\cos(t))) \\
    =&(-\sin t z +\cos t)^{-2B}.
\end{align}
Similarly, the transformations generated by $\check{T}, \check{U}$ are
\begin{align}
\begin{aligned}
    e^{t\check{T}}f(z_0)&=f(z_0+t) \\
    e^{t\check{U}}f(z_0)&=e^{2Bt}f( e^{2t}z_0)= (e^{-t})^{-2B} f( e^{2t}z_0),
\end{aligned}
\end{align}
respectively. Let $\exp(\check{X})$ be a transformation corresponding to $g=\exp(X) \in SL_2(\mathbb{R})$, it acts on a function $f$ on $\mathbb{H}$ as 
\begin{align}
    \check{g}f(z)=\exp(\check{X}) f(z)=j(g,z)^{2B} f(gz).
\end{align}
If $f$ is a automorphic form of weight $-2B \in \mathbb{Z}$ for a Fuchsian $\Gamma$, then  $f$ is invariant under a transformation $\check{\gamma}$ corresponding to $\gamma \in \Gamma$. It is well-known that there exists no automorphic function of weight $m<0$, and an automorphic function of weight $m=0$ is only constant function. So let $f$ be an automorphic form of weight $m=\abs{2B}$. $f$ has zeros in a fundamental domain of $\Gamma$. If $f$ has no zeros, then $1/f$ is an automorphic function of weight $m=-\abs{2B}$. But this statement contradicts with an above fact. 

We assume that $2B\in \mathbb{Z}$. So the magnetic flux through the fundamental domain generated by the $\{4g,4g\}$ tiling is $\phi=4(g-1)\pi B\in 2(g-1) \mathbb{Z}$. Then we can construct a function $\psi_{0,k}$ satisfying
\begin{align}
    \hat{\gamma}^B_j \psi_{0,k} = e^{ik_j}\psi_{0,k}\ (j=1,\cdots ,4),
\end{align}
from an automorphic form $f$ of weight $\abs{2B}$. This condition is a specific case of \eqref{eq:boundary condition of psi_nk}.

We define $u$ by
\begin{align}
    u(z)=\rho(z) \qty(\frac{f(z)}{\abs{f(z)}})^{-\text{sign}B}  ,
\end{align}
where $\rho$ is a function that $\rho$ takes the value $0$ at the zeros of $f$ and satisfies $\rho(\gamma z)=\rho(z)$.
Then we obtain
\begin{align}
    \hat{\gamma}^B_j u(z)=u(z).
\end{align}
With respect to a hyperbolic Bloch state $\psi_k$~\cite{2020arXiv200805489M}, we can write $\psi_{0,k}$ as
\begin{align}
    \psi_{0,k}(z)=\psi_{k}(z)u(z).
\end{align}

\section{Summary}
In this article we considered the hyperbolic band theory in the presence of a magnetic field. We proposed the magnetic Fuchsin group and the magnetic hyperbolic Bloch state. Moreover we explained a relation between Bloch states and automorphic forms. Our formulation is a will open up a new direction of hyperbolic band theory under a magnetic field, which is testable by the modern quantum devices including cQED and quantum simulators. Especially our proposal connects some established theories of algebraic geometry with condensed matter physics in a new way.

\section*{Competing Interests}
The authors declare no conflicts of interest associated with this manuscript.

\section*{Acknowledgements}
K.I. thanks Steven Rayan for useful communications. This work was supported by PIMS Postdoctoral Fellowship Award. Work of KI was supported by the U.S. Department of Energy, Office of Science, National Quantum Information Science Research Centers, Co-design Center for Quantum Advantage (C2QA) under Contract No.DESC0012704. Work of Y.M. was supported by Japan Society for the Promotion of Science (JSPS) Grant-in-Aid for JSPS Research Fellow, No. JP19J20559.  
\bibliographystyle{JHEP}
\bibliography{ref}

\end{document}